\documentstyle[preprint,aps]{revtex}
\input{psfig}
 at 16truept
\font\rmstb=cmbx10 at 16truept
\font\rmft=cmr10 at 14truept
\font\rmftb=cmbx10 at 14truept
\font\rmttb=cmbx10 at 13truept
\font\rmtv=cmr12
 at 12truept
\font\rmbtv=cmbx12
 at 11truept

\begin{document}
\rmft
\hfill THES-TP-96/04
\vskip 1cm
\rmstb
\centerline{SHADOWING OF ULTRAHIGH ENERGY NEUTRINOS}
\vskip 1cm
\rmttb
\centerline{A. NICOLAIDIS and A. TARAMOPOULOS}
\vskip 1cm
\rmft
\centerline{Theoretical Physics Department}
\centerline{University of Thessaloniki,}
\centerline{Thessaloniki 54006, Greece}

\vskip 5cm
\rmftb
\centerline{Abstract}
\rmtv
The rise with energy of the neutrino--nucleon cross section implies that
at energies above few TeV the Earth is becoming opaque to cosmic neutrinos. 
The neutrinos interact with the nucleons through the weak charged current,
resulting into absorption, and the weak neutral current, which provides a 
redistribution of the neutrino energy. We Mellin transform the neutrino
transport equation and find its exact solution in the moment space. A
simple analytical formula is provided, which describes accurately the 
neutrino spectrum, after the neutrinos have traversed the Earth. The 
effect of the weak neutral current is most prominent for an initial 
flat neutrino spectrum and we find that at low energies (around 1 TeV)
the neutrino intensity is even enhanced.
\newpage

Neutrino telescopes will provide a new observational window on our cosmos
and will help us probe the deepest reaches of distant astrophysical 
objects \cite{Gais95}. The enormous energies involved (cosmic neutrinos carry
energy from few TeV to 10$^8$ TeV) might also allow us to enlarge our 
knowledge of particle physics. Since neutrinos interact only weakly,
they travel unhindered by intervening matter and can give us information
about regions of the universe, which are not accessible to traditional
photon astronomy. By the same token, it is extremely difficult to
detect neutrinos and large effective volumes for detection are 
required. Four neutrino observatories are under construction:
DUMAND II in the Pacific Ocean off the coast of Hawaii \cite{Gried93}, NESTOR
in the bay of Pylos, Greece \cite{Resv93}, AMANDA in the ice of Antarctica \cite{Morse93}
and the Baikal NT-200 in the Siberian lake Baikal \cite{Wis93}. Expected neutrino 
sources include atmospheric neutrinos (neutrinos generated in the atmosphere
by the cosmic rays), which dominate other sources at energies below
few TeV, neutrinos generated in active galactic nuclei (AGN) with 
energies up to 10$^6$ TeV and cosmological neutrinos, important
at energies above 10$^6$ TeV \cite{Gais95}.

The neutrino--nucleon cross section rises linearly with energy up
to 10 TeV. At higher energies, due to the propagator effect of the 
intermediate gauge bosons, the rise is slower. At around 40 TeV the
cross section is large enough so that the Earth starts becoming opaque to 
neutrinos. To reduce the background from cosmic rays, at the detection
site we are looking for upward moving muons induced by neutrinos
coming from the other side of the Earth. Therefore, in all relevant
calculations we should include attenuation factors describing
how the neutrino fluxes change as the neutrinos travel through 
the Earth. It is the purpose of the present work to provide an analytic
and accurate representation of the shadowing of highly energetic
neutrinos.

Neutrinos interact with the nucleons through charged and neutral weak 
currents. In the first case a neutrino is transformed into a charged
lepton and we have a neutrino loss, while in the second case the neutrino 
continues along its path with reduced energy. The inclusive neutrino--nucleon
cross section looks generically like
\begin{equation}
{{d^2\sigma } \over { dxdy }} = s_{_B}~{{A(x,Q^2) + (1-y)^2 B(x,Q^2)} \over
{ (1 + s_{_B} x y )^2 }}.
\end{equation}
In the above expression $x$ is the Bjorken scaling variable, $1-y$ is the 
momentum fraction of the produced lepton, $A$ and $B$ are structure functions
expressed in terms of the appropriate parton densities and $s_{_B} = s/M_B^2$
with $M_B$ the mass of the intermediate gauge boson (W or Z). At high energies ($s_{_B} \gg 1$) the propagator term (the term in the denominator) reduces the
rise with energy of the total cross section. Also, at high neutrino energies
the nucleon is probed at exceedingly small $x$ values, where no experimental 
information is available. The neutrino--nucleon cross section at very high
energies is a subject of current research \cite{Frich95,Frich96,Gand79}. For our purposes it is 
sufficient to parametrize the total cross sections in the following
form
\begin{equation}
\sigma_{cc} (E) = a_c \left( { E \over E_0 } \right)^{\beta_c} \label{scc},
\end{equation}
\begin{equation}
\sigma_{nc} (E) = a_n \left( { E \over E_0 } \right)^{\beta_n} \label{snc}.
\end{equation}
Recent calculations \cite{Gand79} provide for neutrino energies above 10$^3$ TeV, 
$\beta_c \simeq \beta_n \simeq \beta \simeq 0.4$, $a_c / a_n \simeq 2.53$. 
An important ingredient in the shadowing process is the $y$ distribution 
of the neutral current cross section. We assume a factorized expression
\begin{equation}
{ {d\sigma_{nc} (E,y)} \over {dy} } = a_n \left( { E \over E_0 }
\right)^{\beta_n} f(y). \label{dsdy}
\end{equation}
There is a mild energy dependence in the $y$ distribution, but for our
practical work we use an $f(y)$ independent of energy. In the domain 
$s_{_B} \gg 1 $ the gauge boson propagator creates a peaking of the cross section
near $y=0$ and we adopt the form \cite{Ber79}
\begin{equation}
f(y) = { 1 \over { \ln\left( 1 / \varepsilon \right) } }~{ 1 \over 
{ \varepsilon + y } }. \label{fy}
\end{equation}
The average $y$ and the parameter $\varepsilon$ are related through 
$<y> = 1 / \ln(1/\varepsilon)$.

The transport equation for neutrinos traversing the Earth is
\begin{equation}
{ {dI(E,\tau)} \over {d\tau} } = - ( \sigma_{cc} + \sigma_{nc} ) 
I(E,\tau ) + { {d\sigma_{nc} } \over {dy} } \otimes I, \label{transport}
\end{equation}
where $d\tau = n(z) dz$ and $n(z)$ is the number density of the 
nucleons encountered by the neutrino along its path through the 
Earth. $I(E,\tau)$ is the intensity of the neutrino flux at
``depth'' $\tau$, with the initial intensity (before entering the
Earth) $\bar{I} (E) = I(E, \tau = 0)$. The convolution is defined by
\begin{equation}
{ {d\sigma_{nc} } \over {dy} } \otimes I = \int { {d\sigma_{nc} (E^\prime , y)}
\over {dy} }~I(E^\prime , \tau ) 
~\delta(E-E^\prime (1-y))~dE^\prime dy. 
\end{equation}
For $n(z)$ we use $n=\rho / M_N $, $\rho$ being the density of the 
Earth. The maximum value $\tau$ can attain is approximately 
$6\times10^{33}$ cm$^{-2}$. To proceed further, we Mellin transform
equ.~(\ref{transport}), using also equs.~(\ref{scc})$-$(\ref{dsdy}). Defining
\begin{equation}
I_k (\tau ) = \int E^k I(E, \tau ) dE,
\end{equation}
we obtain
\begin{equation}
{ {dI_k } \over {d\tau } } = - c_k~I_{k + \beta}, \label{momeq}
\end{equation}
where 
\begin{equation}
c_k = \left( { 1 \over { E_0 } } \right)^\beta ( a_c
+ a_n - a_n f_k ), \label{ck}
\end{equation}
\begin{equation}
f_k = \int_0^1 (1-y)^k f(y) dy.
\end{equation}
For $k=0$, $f_0 =1$ and the total number of neutrinos is reduced because
of the weak charged current, the weak neutral current providing only a
redistribution of energy. $f_k$ is a decreasing function of $k$,
and for sufficiently large values of $k$, $c_k$ corresponds
to absorption created by $\sigma_{tot}$ ( $\sigma_{tot} = \sigma_{cc} 
+ \sigma_{nc}$ ).

Let us define the column vector $J_n$ 
\begin{equation}
J_n = \left( \begin{array}{c} 
I_{0+n\beta} \\
I_{1+n\beta} \\
\vdots \\
I_{k + n\beta} \\
\vdots
\end{array} \right) 
\end{equation}
and the matrix
\begin{equation}
P = \left( \begin{array}{ccccc} 
   0   & R_0     & 0      & 0      & \cdots \\
   0   & 0       & R_1    & 0      & \cdots \\
   0   & 0       & 0      & R_2    & \cdots \\
\vdots & \vdots  & \vdots & \vdots & \nonumber 
\end{array} \right),
\end{equation}
where $R_n$ is the diagonal matrix
\begin{equation}
R_n = - \left( \begin{array}{cccc} 
c_{n\beta} & 0              & 0             & \cdots \\
0          & c_{1+n\beta }  & 0             & \cdots \\
0          & 0              & c_{2+n\beta } & \cdots \\
\vdots     & \vdots         & \vdots        & \nonumber 
\end{array} \right).
\end{equation}
We obtain then 
\begin{equation}
{ d \over {d\tau } } \left( \begin{array}{c}
J_0 \\
J_1 \\
J_2 \\
\vdots \\
J_k \\
\vdots
\end{array} \right) = P \left( \begin{array}{c}
J_0 \\
J_1 \\
J_2 \\
\vdots \\
J_k \\
\vdots
\end{array} \right).
\end{equation}
The formal solution of the above equation is
\begin{equation}
\vec{J} ( \tau ) = \exp(P\tau )~\vec{J} (\tau =0 ).
\end{equation}
In terms of the first component
\begin{equation}
J_0 = \bar{J}_0 + R_0~\bar{J}_1~\tau + { 1 \over {2!} }~R_0~R_1
~\bar{J}_2~\tau^2 + \cdots + { 1 \over {n!} }~R_0~R_1~R_2 \cdots R_{n-1}
~\bar{J}_n~\tau^n + \cdots
\end{equation}
with $\bar{J}_k \equiv J_k (\tau =0 )$. Finally in terms
of the original moments we get
\begin{eqnarray}
I_k (\tau ) = & \bar{I}_k - c_k~\bar{I}_{k + \beta } 
~\tau + { 1 \over {2!} } c_k~c_{k + \beta }~\bar{I}_{k
+ 2\beta}~\tau^2 + \cdots + \nonumber \\
& + (-1)^n {1 \over {n!}} c_k 
~c_{k+2\beta} \cdots c_{k+(n-1)\beta}~\bar{I}_{k+
n\beta}~\tau^n + \cdots. \label{momsol}
\end{eqnarray}
The above expression is the exact solution to equ.~(\ref{momeq}).

The inversion of equ.~(\ref{momsol}) in order to obtain $I(E,\tau )$ is not an easy 
task. We prefer to study limiting behaviors of $I(E, \tau )$ and 
then establish a consistent expansion. It is already stressed that $f_k$
tends to zero as $k$ grows. At sufficiently large $k$, $c_k$
becomes a constant independent of $k$
\begin{equation}
c_k \simeq \left( { 1 \over E_0 } \right)^\beta (a_c + a_n )
\equiv q~~~~~~~~~~~~~~~~({\rm large }~k ). \label{ckq}
\end{equation}
Substituting the above expression into equ.~(\ref{momsol}) we get
\begin{equation}
I_k (\tau ) \simeq \int \bar{I} (E) E^k \exp\left[
-\sigma_{tot} (E) \tau \right] dE.
\end{equation}
Therefore, at large energies (large $k$ ) the neutrino intensity
is well represented by the standard absorption formula
\begin{equation}
I_{abs} (E,\tau ) = \bar{I} (E) \exp\left[ -\sigma_{tot} (E) \tau 
\right]. \label{iabs}
\end{equation}
A systematic approach emerges, where the coefficients $c_k$
for $k$ larger than some value are replaced by $q$ (equ.~(\ref{ckq})) and
for the others we use the exact expression (equ.~(\ref{ck})). The first term in 
our expansion is obtained by keeping $c_k$ in equ.~(\ref{momsol}) and
replacing all $c_{k + n\beta } (n \geq 1 ) $ by $q$. We find
\begin{equation}
I_k (\tau ) \simeq \sum_{n=0}^{\infty} \bar{I}_{k + n\beta} 
{ { \left( -q \tau \right)^n } \over {n!} } + { {a_n} \over 
{a_c + a_n} } f_k \left[ \bar{I}_k - 
\sum_{n=0}^{\infty} \bar{I}_{k +n\beta } { { \left( -q \tau \right)^n } 
\over {n!} } \right].
\end{equation}
Returning to the energy variable we obtain
\begin{equation}
I_1(E,\tau )= I_{abs} (E,\tau ) + { {a_n} \over {a_c + a_n}} 
f(y) \otimes \left[ \bar{I} - \bar{I}_{abs} \right]. \label{appr1}
\end{equation}
Under the assumption of a power law initial neutrino spectrum
\begin{equation}
\bar{I} (E) = \bar{I} (E_0) \left( { E \over {E_0 }}\right)^{-\gamma} \label{inispec}
\end{equation}
equ.~(\ref{appr1}) provides
\begin{eqnarray}
I_1 (E, \tau ) = & \bar{I} (E) \exp\left[ -\sigma_{tot} (E) \tau 
\right] +~~~~~~~~~~~~~~~~~~~~~~~~~~~~~~~~~~~~~~~~~~~~~~~~~~~~ \nonumber \\
& + \bar{I} (E)~{ {\sigma_{nc} (E) } \over { \sigma_{tot} (E) } }~~
{\displaystyle \int_0^1} dy~f(y) (1-y)^{\gamma -1 } 
\left[ 1 - \exp\left(
- \left( { 1 \over {1-y}} \right)^\beta \sigma_{tot}(E) \tau \right) 
\right]. \label{appr1b}
\end{eqnarray}
Successive terms in our expansion can be obtained by using the exact
expression for more than one $c_k$ (two, three, $\cdots$).

One could treat the second term in equ.~(\ref{transport}) as a source function $F$. 
Defining
\begin{equation}
F(E,\tau ) = \int_0^1 { {d\sigma_{nc}}
\over {dy}} ( { _E \over ^{1-y}},~ _y )~I( { _E \over ^{1-y}},~ _\tau )
~{ {dy} \over {1-y} } \label{fdef}
\end{equation}
and multiplying equ.~(\ref{transport}) by $\exp[\sigma_{tot}(E) \tau ]$ results in
\begin{equation}
{ {dH} \over {d\tau } } = \exp\left( \sigma_{tot} \tau \right) F ,
\end{equation}
where $H=\exp(\sigma_{tot} \tau ) I$. Direct integration yields
\begin{equation}
I_s (E,\tau ) = \bar{I} (E) \exp\left( -\sigma_{tot} \tau \right)
+ \exp\left( -\sigma_{tot} \tau \right) \int_0^\tau F(E,t)
\exp\left( \sigma_{tot} t\right) dt. \label{fsol}
\end{equation}
Reasonable approximations may be obtained by inserting ``appropriate''
forms for $I$ in equ.~(\ref{fdef}). Assuming the generic form
\begin{equation}
I_p (E,\tau ) = \bar{I} (E) \exp\left[ -\sigma_p (E) \tau \right] \label{genf}
\end{equation}
with $\sigma_p (E)$ to be specified, equ.~(\ref{fsol}) provides
\begin{eqnarray}
I_s (E,\tau ) = & \bar{I} (E) \exp \left[ -\sigma_{tot} \tau \right] 
+ ~~~~~~~~~~~~~~~~~~~~~~~~~~~~~~~~~~~~~~~~~~~~~~~~~~~~~~~~~~\nonumber \\
& + \bar{I} (E)~{ {\sigma_{nc} (E) } \over {\sigma_{tot} (E) } }~~
{\displaystyle \int_0^1} dy~f(y) (1-y)^{\gamma -1 }~~ 
{ { \exp\left[ -\sigma_{tot}(E)
\tau \right] - \exp\left[ - \left( { 1 \over {1-y} }\right)^\beta
\sigma_p (E) \tau \right] } \over { \left[ { {\sigma_p (E) } \over
{\sigma_{tot} (E) } } - \left( 1-y \right)^\beta \right] } }. \label{appr2}
\end{eqnarray}
Adopting $\sigma_p = 0$ (i.e. $I_p = \bar{I} (E) $ ) we obtain an
expression similar to equ.~(\ref{appr1b}). Both expressions are valid for short $\tau$ 
and they overestimate significantly the actual result at large $\tau$. 
The other obvious choice is $\sigma_p (E) = \sigma_{tot} (E) $, i.e.
we insert in equ.~(\ref{fdef}) the zeroth-order answer for $I(E,\tau )$.
We proceeded to a numerical solution of equ.~(\ref{transport}) and in fig.~1 we show 
the shadowing factor defined by
\begin{equation}
S (E, \tau ) = { {I(E, \tau ) } \over { \bar{I} (E) } } 
\end{equation}
as a function of the energy for the maximum value of $\tau$. For the
initial neutrino spectrum we used the form of equ.~(\ref{inispec}) with 
$\gamma = 2$. Existing calculations of the unresolved AGN neutrino
flux \cite{Stec9192,Nel93,Kaz93} give differential spectra with spectral index $\gamma$
varying between 0 and 2 over the energy range. The absorption 
factor suggested by equ.~(\ref{iabs}) is shown also in fig.~1 (long dashed curve). It
remains always below the numerical shadowing factor (solid line)
and the two curves approach each other at large energies, as expected. 
The dash-dotted curve represents the shadowing factor implied by equ.~(\ref{appr2})
with $\sigma_p (E) = \sigma_{tot} (E)$. It is a significant improvement
over the simple absorption formula (equ.~(\ref{iabs})), although it lags 
behind the numerical result. 

Searching a more ``appropriate'' form for the neutrino intensity, we notice 
that for the specific form of $f(y)$ (equ.~(\ref{fy})), we have
\begin{equation}
f_k = 1 - { 1 \over {\ln\left( {1\over \varepsilon} \right) } } 
\left( 1 + {1\over 2} + {1\over 3} + \cdots + {1\over k } \right).
\end{equation}
Then for moderate values of $k$ ($k \geq 1$, but not very
large) $c_k$ remains relatively constant at
\begin{equation}
c_k \simeq \left( { 1 \over {E_0} }\right)^\beta \left(
a_c + <y> a_n \right).
\end{equation}
With the above $c_k$, the neutrino intensity is provided by an 
absorption formula, equ.~(\ref{genf}), with 
\begin{equation}
\sigma_p (E) = \sigma_{cc} (E) + <y> \sigma_{nc} (E). \label{spy}
\end{equation}
The shadowing factor given by equs.~(\ref{appr2}) and (\ref{spy}) is shown in fig.~1
(short dashed curve) and it is in excellent agreement with the numerical 
simulation.

Another approximate solution to the neutrino transport equation has been 
suggested in ref \cite{Ber86}. Defining the effective cross section
\begin{equation}
\sigma_{eff} (E) = \sigma_{cc} (E) + \sigma_{nc} (E) -
\sigma_{nc} (E) \int_0^1 dy (1-y)^{\gamma - \beta -1 } f(y), \label{apprr}
\end{equation}
the neutrino ``regeneration'' effect is (incorrectly) exponentiated through
the formula 
\begin{equation}
I_R (E,\tau ) = \bar{I} (E) \exp\left[ -\sigma_{eff} (E) \tau \right]. \label{seff}
\end{equation}
The ratio of the suggested shadowing factor by equ.~(\ref{seff}) to the 
numerical one is shown in the inset of fig.~1 (dashed line). Clearly
equ.~(\ref{seff}) overestimates the neutral current contribution. In the 
same inset we show the corresponding ratio for our formula,
equs.~(\ref{appr2}) and (\ref{spy}) (solid line).

The neutral current removes neutrinos from the large energy part
of the spectrum and populates the lower energy part of the spectrum. 
It is evident that the shadowing factor depends upon the initial
spectrum, since a steeply falling spectrum does not provide considerable 
regeneration, while a spectrum extending to high energies (i.e. 
flat spectrum) induces large regeneration. 
Calculations of the AGN neutrino flux \cite{Stec9192} suggest a flat energy 
distribution ($\gamma = 0$) up to energies $E_{max} = 10^3$ TeV. 
Fig.~2 presents the shadowing factor for $\gamma = 0$ with the other
assumptions unchanged as in fig.~1. The disagreement of the shadowing
factor implied by equ.~(\ref{seff}) to the numerical calculation (dashed
line in the inset) increases. Fig.~2 shows that, contrary to simple
absorption, the shadowing factor at low energies can become greater than 
1.0. This  is an effect we could anticipate. Neutral current shifts 
the neutrino energy to lower values. If the new energy is low enough, such
that $\sigma_{tot} (E) \tau \ll 1$, then the neutrino traverses the
Earth unabsorbed. Therefore, at such nominal energy, apart from the initial
neutrinos, we will have the neutrinos displaced in energy by the
neutral current. In our calculations (fig.~2) the shadowing factor at
1 TeV is 1.36.

The use of ultrahigh energy neutrinos in order to image the Earth's
internal structure \cite{Vol74,Ruj83,Reno88,Berg92,Kuo95,Mann95} has been considered for some time.
It is highly interesting to obtain further information on the density 
distribution of the Earth, independently of the seismological determinations. 
In these investigations the modified neutrino flux is given by simple 
absorption, ignoring the neutrino regeneration by the neutral current. We 
provided a simple, analytical and accurate description, equs.~(\ref{appr2}) and (\ref{spy}),
for the neutrino propagation inside the Earth, for all the values of
energy $E$ and depth $\tau$. Our formalism will be essential for
all reliable evaluations of the event rates in the neutrino telescopes
under construction \cite{Gried93,Resv93,Morse93,Wis93}. In the present work we used simple
parametrizations for the neutrino cross sections, which may not be
very accurate. Our main concern here is to establish a correct 
procedure, rather than to provide detailed numbers. At very high energy,
the cross sections sample the nucleon parton densities at small $x$, 
a kinematical region where new physical phenomena (BFKL physics \cite{Grib8394}) are
operative. The shadowing factor is very sensitive to the actual 
magnitude of the cross section and therefore the neutrino passage
through the Earth will provide information about the structure of
the nucleon at small $x$. It seems that the detection of ultrahigh energy 
cosmic neutrinos will be important not only for astrophysics, but also for 
particle physics and geophysics. 
\vskip 0.3cm
\rmbtv
\noindent Acknowledgements.
\vskip 0.3cm
\rmtv
One of us (A.N.) would like to acknowledge useful
discussions with P. Gondolo, V. Stenger, J. Learned, L. Resvanis, I. 
Sarcevic, F. Stecker, A. Tarantola, D. Kazanas and S. Ichtiaroglou. This 
research was supported in part by the Fulbright Foundation and the EU program 
``Human Capital and Mobility''. 

\begin{figure}
\psfig{figure=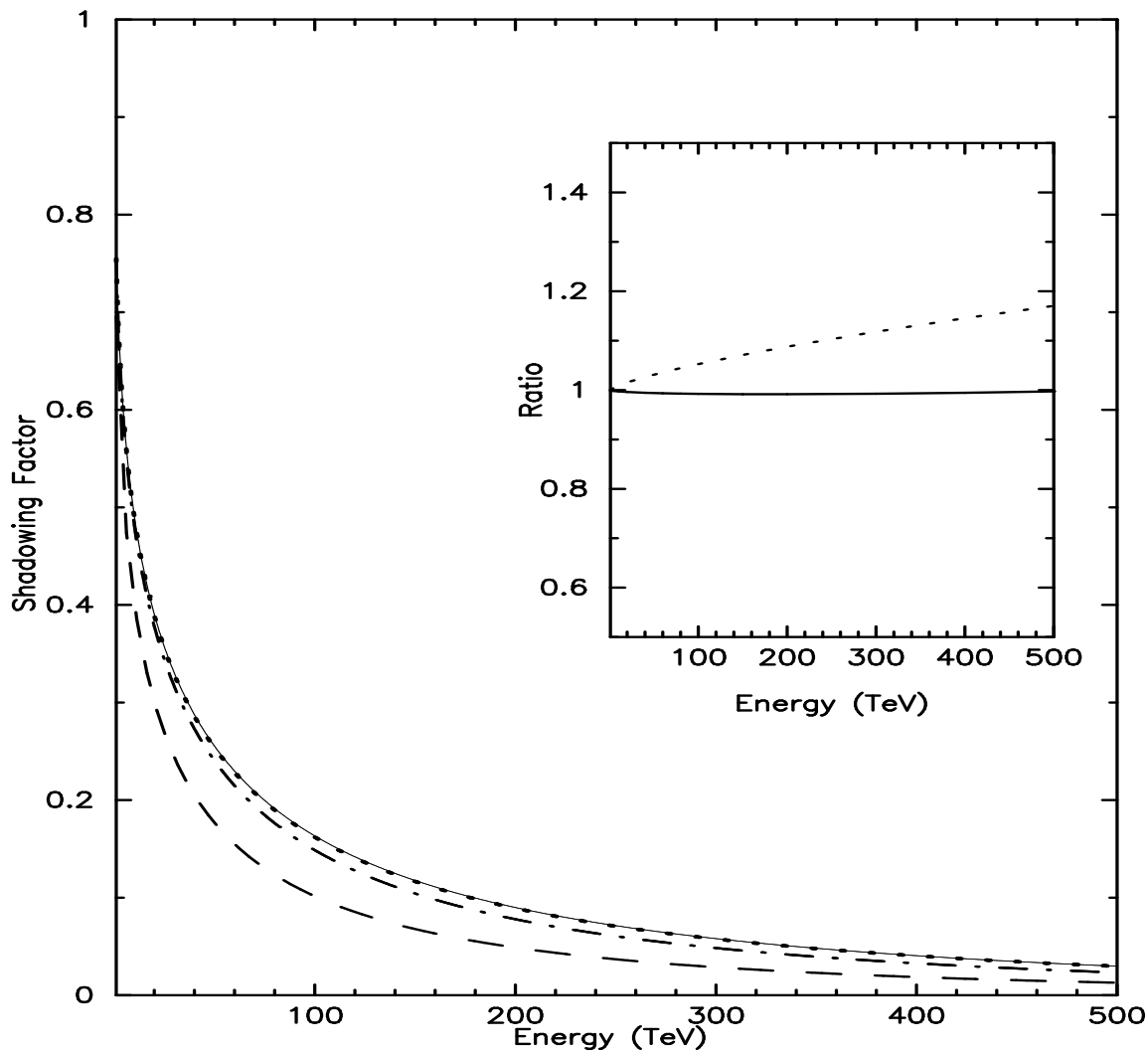,width=6in,height=5.5in}
\vskip 0.3in
\caption{ The shadowing factor for neutrinos traveling along the Earth's
diameter, as a function of the energy for spectral index $\gamma = 2.0$.
The long-dashed curve is the absorption factor given by equ.(21), the 
dot--dashed curve is the shadowing factor implied by equ.(30) with $\sigma_p =
\sigma_{tot}$, the short--dashed curve is the shadowing factor implied
by equs.(30) and (34), while the solid line represents the numerical
evaluation of the shadowing factor. The inset shows the ratio
of the proposed shadowing factor to the numerical one. The dashed--curve is 
derived from equ.(36) and the solid curve is derived from equs.(30) and 
(34).}
\label{g2}
\end{figure}

\begin{figure}
\psfig{figure=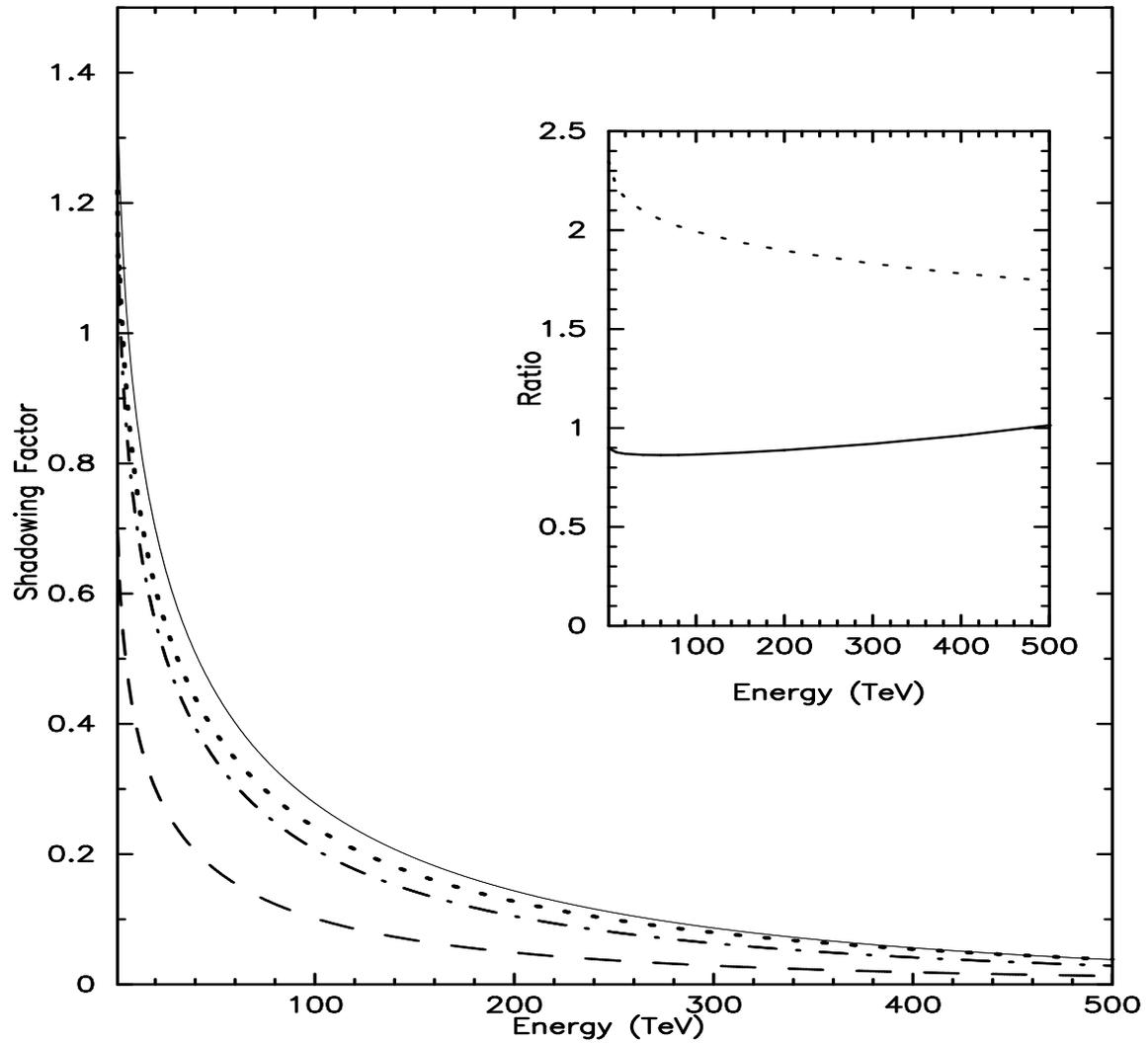,width=6in,height=5.5in}
\vskip 0.3in
\caption{ The shadowing factor for neutrinos traveling along the 
Earth's diameter, as a function of the energy, for spectral index $\gamma = 0$ 
and $E \leq 1100$ TeV. Different curves correspond to the same conditions as
in fig. 1.}
\label{g0}
\end{figure}

\end{document}